\documentclass[aps,prb,tightenlines,showpacs,twocolumn,superscriptaddress]{revtex4-1}  %  

\usepackage{amsmath,amssymb,amsfonts,bm}
\usepackage{graphicx}
\usepackage{dcolumn}
\usepackage{color}
\usepackage[colorlinks=true,linkcolor=blue,citecolor=blue]{hyperref}% 

\begin{document}

%-----------------------------------------------------------------
% documentation    title,  authors,   abstract,   pacs
%-----------------------------------------------------------------

\title{Kondo temperature of Anderson impurity model in a quantum wire with spin-orbit coupling}
\author{Liang Chen}
\email[Corresponding Email: ]{slchern@ncepu.edu.cn}
\affiliation{Mathematics and Physics Department, North China Electric Power University,Beijing, 102206, China}
%\affiliation{Beijing Computational Science Research Centre, Beijing, 100084, China}
%\author{Bo Hu}
%\affiliation{Mathematics and Physics Department, North China Electric Power University,Beijing, 102206, China}
\author{Rong-Sheng Han}
\email[Corresponding Email: ]{hrs@ncepu.edu.cn}
\affiliation{Mathematics and Physics Department, North China Electric Power University,Beijing, 102206, China}
\date{\today}

\begin{abstract}
We use two different methods, the Hirsch-Fye quantum Monte Carlo simulation and the slave-boson mean field analysis to estimate the effect of spin-orbit coupling on the Kondo temperature in a quantum wire. In quantum Monte Carlo simulation, we calculate the product of spin susceptibility and temperature for different spin-orbit couplings and impurity energies. The variation of the Kondo temperature is estimated via the low temperature universal curves. In the mean field analysis, the Kondo temperature is estimated as the condensation temperature of slave-boson. Both the two methods demonstrate that the Kondo temperature is almost a linear function of the spin-orbit energy, and that the Kondo temperature is suppressed by the spin-orbit coupling. Our results are dramatically different from those given by the perturbative renormalization group analysis. 

\end{abstract}

\pacs{72.10.Fk, 71.70.Ej, 72.15.Qm}

\maketitle

%-----------------------------------------------------------------
% The body of the paper
%-----------------------------------------------------------------

\section{Introduction} \label{sec1} %-----------------------------
Spin-orbit coupling \cite{RolandWinkler2003} (SOC), the interaction between a quantum particle's spin and the effective magnetic field induced by its (orbital) motion, has attracted much attention as an useful tool for the realization of spintronics in the last few decades \cite{RevModPhys.76.323,PhysRevLett.89.046801,PhysRevB.84.153402}. In recent years, SOC plays an important role in the topological nontrivial quantum materials, e.g., topological insulator \cite{RevModPhys.82.3045,RevModPhys.83.1057}, topological superconductor \cite{PhysRevLett.105.077001}, quantum anomalous Hall effect \cite{Chang167}, Weyl semimetal \cite{PhysRevB.83.205101}, Majorana fermion \cite{Mourik1003,Das2012}, etc. Magnetic doping in these systems is an efficient method to tune and control the transport properties for some of these systems, e.g. breaking the time-reversal symmetry and inducing a finite band gap on the surface of topological insulator. 

The Kondo effect in some of these SOC systems has been theoretically investigated \cite{PhysRevB.87.195122} and experimentally confirmed \cite{nanoLett2010Cha}. However, as a fundamental problem, the effect of SOC on the Kondo temperature is still unclear as far as we know. In recent years, several theoretical works have investigated this problem\cite{Malecki2007,PhysRevLett.108.046601,PhysRevB.84.193411,PhysRevB.85.081107,PhysRevB.93.075148,PhysRevB.93.241111,jpcm2016chen}. By studying the standard Kondo model in a two-dimensional electron gas, Malecki \cite{Malecki2007} conclude that the Kondo effect is robust against Rashba SOC. However, it is demonstrated \cite{PhysRevLett.108.046601,PhysRevB.94.125115} that the standard Kondo model in SOC systems is incomplete. Under a generalized Schrieffer-Wolff transformation of the Anderson impurity model in two-dimensional electron gas with Rashba SOC, Zarea et al. \cite{PhysRevLett.108.046601} find that, in addition to the usual Kondo interaction, these is another Dzyaloshinskii-Moriya interaction between the magnetic impurity and the conduction electrons. Using the perturbative renormalization group analysis, they conclude that the Kondo temperature is exponentially enhanced by the SOC. However, both the numerical renormalization group calculation \cite{PhysRevB.84.193411,PhysRevB.93.075148} and the quantum Monte Carlo simulation \cite{jpcm2016chen} show that the Kondo temperature is almost a linear function of the spin-orbit energy. 

Recently, Sousa {et al.} \cite{PhysRevB.94.125115} find that the effect of SOC on the Kondo temperature in the one-dimensional quantum wire is special due to the discontinuity of the fermi surface. In addition to the traditional spin-flipping scattering and the Dzyaloshinskii-Moriya interaction in SOC system, there is another process for one-dimensional quantum wire, the Elliott-Yafet scattering. The combination of these two processes, Dzyaloshinskii-Moriya interaction and Elliott-Yafet scattering, makes the Kondo temperature exponentially enhanced by the SOC. However, these conclusions are based on the perturbative renormalization group analysis, a non-perturbation study is still opening. In this paper, we use the Hirsch-Fye quantum Monte Carlo (HFQMC) simulation and the slave-boson mean field theory (SBMF) to study the single-impurity Anderson model in quantum wire with both Rashba and linear Dresselhaus SOC, and estimate the effect of the SOC on the Kondo temperature. 

The rest of the paper is organized as follows. We present the Anderson impurity model in Sec. \ref{sec2}. The HFQMC simulations are given in Sec. \ref{sec3}. Sec. \ref{sec4} presents the formulation and results of the SBMF analysis. A discussion is given in Sec. \ref{sec5}. 

%\begin{figure}[t]
%	\includegraphics[width=\columnwidth]{fig1.eps} 
%	\caption{(color online) (a) Schematic representation of the model. A quantum dot hybridized to a quantum wire with spin-orbit coupling. The wire is set to lie along the $z$-axis, $V_k$ is the strength of the hybridization. (b) A typical band structure of the quantum wire with spin-orbit coupling. The spin-degeneracy is split by the spin-orbit coupling. (c) A typical band structure of the quantum wire with both spin-orbit coupling and applied magnetic field. The two-fold degeneracy at the Kramer point is broken by applied magnetic field. }%
%	\label{fig1}%
%\end{figure}

\section{Model Hamiltonian}\label{sec2} %-------------------------
The single-impurity Anderson model in quantum wire with Rashba and linear Dresselhaus SOC is formulated as, $H=H_{\text{wire}}+H_{\text{dot}}+H_{\text{hyb}}$, where, 
\begin{equation}
H_{\text{dot}}=\sum_{s=\uparrow,\downarrow}\varepsilon_d n_s + Un_{\uparrow}n_{\downarrow},\label{eq1}
\end{equation}
describes the quantum impurity, in which $s=\uparrow,\downarrow$ refer to the spin-up and spin-down impurity states, $n_{s}=d_{s}^{\dagger}d_s$ is the occupation number of spin-s state, $d_s^{\dagger}$ and $d_s$ are the creation and annihilation operators of the spin-s impurity state, respectively.  $U$ is the Hubbard interaction between the occupied spin-up and spin-down states on the impurity energy level $\varepsilon_d$. The Hamiltonian of the quantum wire is given by, 
\begin{equation}
	H_{\text{wire}}=\sum_{k,ss'}\left[\varepsilon_{k}+k\left(\gamma_{D}\sigma_{ss'}^{x}-\gamma_{R}\sigma_{ss'}^{y}\right)\right]c_{ks}^{\dagger}c_{ks'},\label{eq2}
\end{equation}
where $k$ is the wave-vector of the conduction electron state along the $z$ axis, $\varepsilon_k=\hbar^2k^2/2m^*-E_F$ is the kinetic energy represented with effective mass $m^*$. $E_F$ is the Fermi energy. $\gamma_R$ and $\gamma_D$ are the Rashba and the linear Dresselhaus SOC. $\sigma^{(x,y)}$ are the first two Pauli matrices. $c_{ks}^{\dagger}$ and $c_{ks}$ are the creation and annihilation operators of the electron state with wave-vector $k$ and spin $s$. The hybridization between the isolated impurity and the quantum wire is represented as, 
\begin{equation}
	H_{\text{hyb}}=\sum_{ks}\left(V_kc_{ks}^{\dagger}d_s+V_k^*d_{s}^{\dagger}c_{ks}\right), \label{eq3}
\end{equation}
where $V_k$ is the hybridization matrix element. Without loss of generality, we assume that the hybridization is short ranged, so that $V_k$ is wave-vector independent, $V_k=V$. 

The conduction band electron states in the quantum wire can be integrated out to get the partition function of the impurity states, 
\begin{align}
	\mathcal{Z}&=\int{\mathcal{D}d_s^{\dagger}\mathcal{D}d_s}e^{-\mathcal{S}}, \label{eq4} \\
	\mathcal{S}&=\int_{0}^{\beta}\mathrm{d}\tau\left[d_s^{\dagger}\left(\frac{\partial}{\partial\tau}+\varepsilon_d+\Sigma_d\right)d_s+Ud_{\uparrow}^{\dagger}d_{\uparrow}d_{\downarrow}^{\dagger}d_{\downarrow}\right], \label{eq5}
\end{align}
where $\beta=1/k_BT$ is the inverse temperature, $k_B$ is the Boltzmann constant. The effect of the conduction band electron states is included in the self-energy of the impurity states, $\Sigma_d$. In the imaginary-frequency representation, the self-energy is given in the following form, 
\begin{equation}
	\Sigma_d(i\omega_n)=-\frac{i~\Gamma~\text{sgn}[\text{Im}(i\omega)]}{\sqrt{2(i\omega_n+E_F)+E_{\gamma}}},  \label{eq6}
\end{equation}
where $\Gamma=\rho{V^2}$ is the hybridization strength in the conventional form, $\rho$ is the density of states (DOS). $\omega_n=(2n+1)\pi{k_BT}$ is the Matsubara frequency of Fermion. It is easy to check that, the total DOS of the non-interacting quantum wire, $\rho(\varepsilon)=\rho_{\uparrow}(\varepsilon)+\rho_{\downarrow}(\varepsilon)$, is proportional to the imaginary part of $\Sigma_d(\varepsilon+i0^+)$. In the non-interacting limit, $U\rightarrow0$, the imaginary-time free Green's function of the impurity state is given by the Fourier transformation, 
\begin{equation}
	G_0(\tau)=\frac{1}{\beta}\sum_{n}e^{-i\omega_n\tau}\left[i\omega_n+E_F-\varepsilon_d-\Sigma(i\omega_n)\right]^{-1}. \label{eq7}
\end{equation}
Once we get the imaginary-time free Green's function, we can use the HFQMC method to solve the magnetic impurity problem. 

\begin{figure}[t]
	\includegraphics[width=0.8\columnwidth]{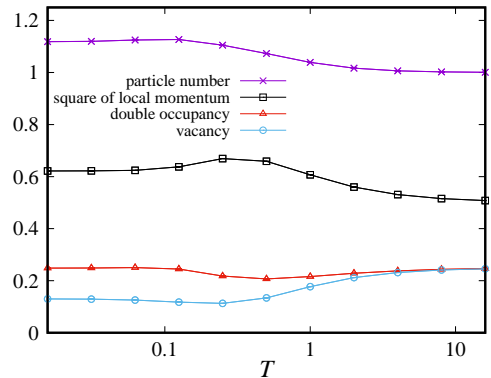} 
	\caption{(color online) The particle number $n=n_{\uparrow}+n_{\downarrow}$, square of local moment, $m^2=(n_{\uparrow}-n_{\downarrow})^2$, vacancy $(1-n_{\uparrow})(1-n_{\downarrow})$, and double occupancy (with error-bars) vs temperature. Parameters used in the simulation: hybridization $\Gamma=0.05\pi$, Fermi energy $E_F=0.2$, impurity energy level $\varepsilon_d=-0.5$, Hubbard interaction $U=1$, spin-orbit energy $E_{\gamma}=0$.}%
	\label{fig1}%
\end{figure}

\begin{figure*}[tbh]
	\includegraphics[width=0.8\textwidth]{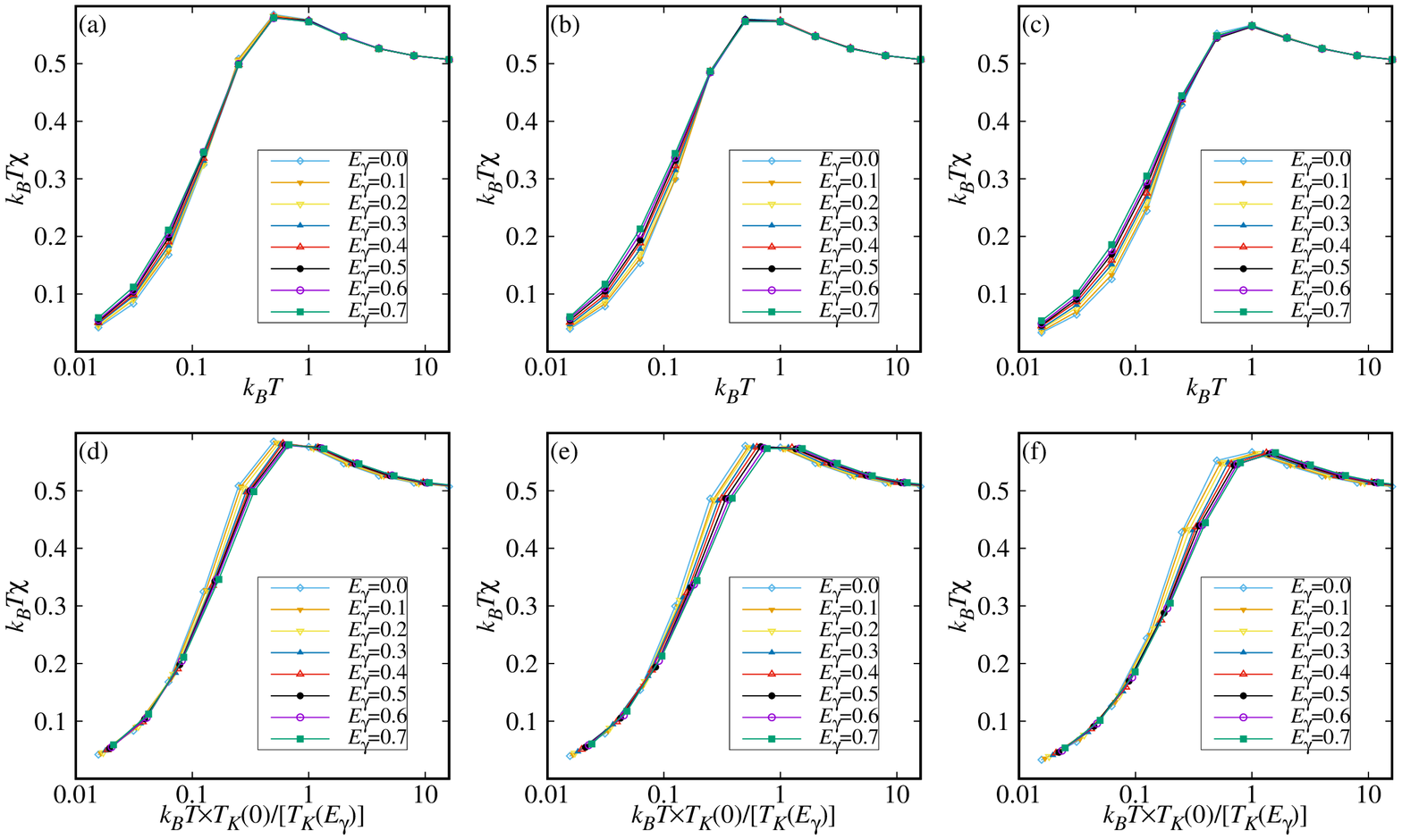} 
	\caption{(color online) (a)-(c), the product $k_BT\chi$ vs temperature for different impurity energies, spin-orbit energies: (a) $\varepsilon_d=-0.3$, (b) $\varepsilon_d=-0.5$, (c) $\varepsilon_d=-0.7$. (d)-(f) the product $k_BT\chi$ vs rescaled temperatures corresponding to (a)-(c), respectively. Model parameters used in these simulations: $\Gamma=0.05\pi$, $U=1.0$, $E_F=0.2$, $h=0$.}%
	\label{fig2}%
\end{figure*}

\begin{figure}[bt]
	\includegraphics[width=\columnwidth]{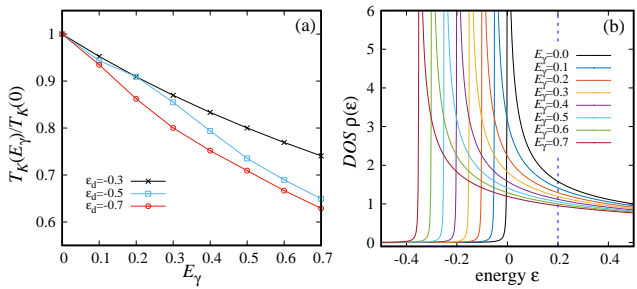} 
	\caption{(color online)(a) The variation of Kondo temperature $T_K$ vs spin-orbit energy $E_{\gamma}$ for three different impurity energies $\varepsilon=-0.3$, $-0.5$, $-0.7$, fitted from Fig. \ref{fig2}. (b) The DOS for different spin-orbit energies from $E_{\gamma}=0$ to $E_{\gamma}=0.7$.}%
	\label{fig3}%
\end{figure}

\section{Hirsch-Fye Quantum Monte Carlo simulation}\label{sec3} %----------------------------
In the HFQMC calculation, the Hubbard interaction is simulated by the statistical average over the auxiliary Ising pseudo-spin configurations. In the following numerical simulations, all of the results are averaged over 20 independent samples, and every sample runs over 2000 HFQMC loops after warm-up. HFQMC is numerical non-perturbative method, it is widely used in the previous works, i.e. the local moment of magnetic impurity\cite{PhysRevLett.56.2521,PhysRevB.38.433}, the competition between Kondo screen and Ruderman-Kittel-Kasuya-Yosida (RKKY) interaction for two-impurity Anderson model \cite{PhysRevB.35.4901,PhysRevB.40.4780}, the local moment formation in dilute magnetic semiconductors \cite{PhysRevB.76.045220} and graphene \cite{PhysRevB.84.075414}, the RKKY interaction in a topological insulator \cite{PhysRevB.89.115101}.

Fig \ref{fig1} shows four important static physical quantities, the particle number $n=n_{\uparrow}+n_{\downarrow}$, the square of local moment $m^2=(n_{\uparrow}-n_{\downarrow})^2$, the vacancy $(1-n_{\uparrow})(1-n_{\downarrow})$, and the double occupancy $n_{\uparrow}n_{\downarrow}$ vs temperature. From this plotting, we find three results. Firstly, the particle number $n=n_{\uparrow}+n_{\downarrow}\approx1$, which demonstrates that the Hubbard interaction $U=1$ is strong enough for the formation of local moment states. Secondly, the square of local moment, $m^2=(n_{\uparrow}-n_{\downarrow})^2$, has a maximum point at the intermedia region, $0.1<T<1$, which demonstrates that the local moment is formed in this region and further screened in the strong correlation region, $T<0.1$. Thirdly, the statistical error bars are plotted and most of them are too small to be observed, which demonstrates that 20 samples and 2000 loops for each sample are accurate enough for this problem.  

Now we consider the effect of SOC on the Kondo temperature. This situation has been studied in Ref. [\onlinecite{PhysRevB.94.125115}] by using the perturbative renormalization group method. Here we use the universal curve to study the problem. In the seminal paper [\onlinecite{PhysRevB.21.1003}], the Kondo temperate is related to the product of temperature and impurity spin susceptibility via the following universal relationship, 
\begin{equation}
	\Phi[4k_BT\chi(T)-1]=\text{ln}(T/T_K), \label{eq8}
\end{equation}
where $\Phi$ is a universal function and the impurity spin susceptibility $\chi(T)$ is defined as, 
\begin{equation}
	\chi(T)=\int_0^{\beta}\mathrm{d}\tau\langle{S_{z}(\tau)S_z(0)}\rangle, \label{eq9} 
\end{equation}
$S_z(\tau)=[n_{\uparrow}(\tau)-n_{\downarrow}(\tau)]/2$. Fig. \ref{fig2} (a)-(c) show the product of temperature and spin susceptibility vs temperature for different parameters. According to Eq. (\ref{eq8}), we keep the Y-axis unchanged and the X-axis rescaled, such that in the low temperature region ($k_BT<0.1$) the curves coincide to each other. The results are plotted in Fig. \ref{fig2} (d)-(f). The rescaling defines the variation of the Kondo temperature $T_K(E_{\gamma})/T_K(0)$ vs the spin-orbit energy, as shown in Fig. \ref{fig3}(a). One can find that (a) the Kondo temperature $T_K$ is almost a linear function of the spin-orbit energy $E_{\gamma}$, (b) the Kondo temperature is suppressed by the spin-orbit energy. This is qualitatively different from those given by the perturbative renormalization group theory, which show that the Kondo temperature is exponentially enhanced by the spin-orbit energy. Qualitatively, this is easy to understood. Fig. \ref{fig3}(b) shows the DOS for different spin-orbit energies. One can find that, the larger spin-orbit energy will induce a smaller DOS on the fermi surface, $E_F=0.2$, which demonstrates that a lower Kondo  temperature is required to screen the local moment state. For the model (\ref{eq1})-(\ref{eq3}), our non-perturbative simulations show that  the variation of the Kondo temperature is dominated by the alternation of the DOS on the Fermi surface, rather than the Dzyaloshinskii-Moriya interaction or Elliott-Yafet spin-flip scattering mechanism proposed in Ref. [\onlinecite{PhysRevB.94.125115}].

\section{Slave-boson mean field analysis}\label{sec4}

\begin{figure*}[tb]
	\includegraphics[width=\textwidth]{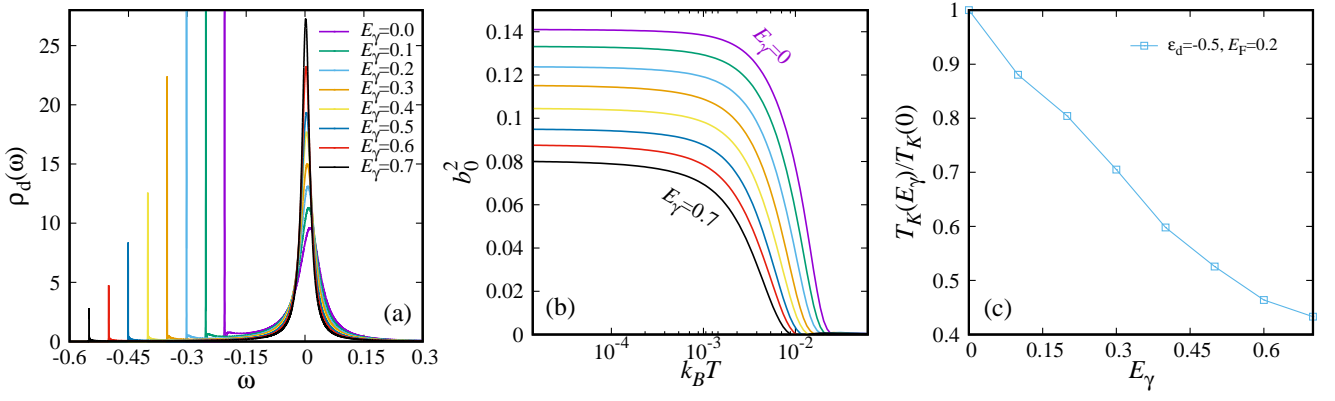} 
	\caption{(color online) (a) The density of states of the impurity for different spin-orbit energies from $E_{\gamma}=0$ to $E_{\gamma}=0.7$. (b) $b_0^2$ vs temperature for different spin-orbit energies. (c) The variation of the Kondo temperature vs spin-orbit energy estimated from the vanishing point of $b_0^2$. The other parameters are chosen as $E_F=0.2$, $\Gamma=0.05\pi$.}%
	\label{fig4}%
\end{figure*}

In order to be more persuasive, we use another approach, the SBMF method, to estimate the effect of the SOC on the Kondo temperature. SBMF is used in the previous studies of the Anderson impurity in the mixed-valence regime \cite{PhysRevB.29.3035,PhysRevB.35.5072}, the Anderson lattice model \cite{PhysRevB.47.5095,PhysRevB.66.045112,NUNES2006313}, and the low-energy physics in strong correlation systems such as the magnetic impurity resonance states in a $d$-wave superconductor \cite{PhysRevB.63.020506} and iron-based superconductors \cite{PhysRevB.81.014524}. We consider a strong on-site Coulomb interaction on the impurity, $U\rightarrow\infty$. In this limit, only the empty state and the singly occupied states  are allowed. An auxiliary boson field, is introduced to reformulate the annihilation and creation operators of the impurity state as, 
\begin{equation}
d_s=b^{\dagger}f_s,~~~~~d_s^{\dagger}=f_s^{\dagger}b,  \label{eq10}
\end{equation}
where $b^{\dagger}$ and $b$ are the boson creation and annihilation operators of the empty states, $f_{s}^{\dagger}$ and $f_s$ are the fermion operators of the singly occupied states. These states expand the whole Hilbert space of the impurity with strong Coulomb interaction, such that these operators have to obey the local constraint $b^{\dagger}b+\sum_{s}f_{s}^{\dagger}f_s=1$.  Under the mean field approximation, the local constraint is approximated by adding a term $\lambda_0(b^{\dagger}b+\sum_{s}f_{s}^{\dagger}f_s-1)$ to the model Hamiltonian, where $\lambda_0$ is a Lagrangian multiplier. The boson operators, $b$ and $b^{\dagger}$ are approximated by a complex number $b_0$ and its complex conjugate $b_0^{*}$. Without loss of generality, here we assume that $b_0$ is a real number. Substituting Eq. (\ref{eq10}) in to Eq. (\ref{eq1}) and Eq. (\ref{eq3}), we get, 
\begin{align}
	H_{\text{dot}}&=\sum_{s}\tilde{\varepsilon}_df_s^{\dagger}f_s+\lambda_0(b_0^2-1), \label{eq11} \\
	H_{\text{hyb}}&=\sum_{k,s}\left(\tilde{V}_{k}c_{k,s}^{\dagger}f_s+\tilde{V}_{k}^{*}f_s^{\dagger}c_{k,s}\right), \label{eq12}
\end{align}
in the SBMF approximation, where $\tilde{\varepsilon}_d=\varepsilon_d+\lambda_0$ is the renormalized impurity energy and $\tilde{V}_{k}=b_0V_k$ is the renormalized hybridization strength. Following the procedure presented in Sec. \ref{sec3}, we can finish the integration over the fermion states and get the partition function, 
\begin{equation}
	\mathcal{Z}=\exp\left\{\beta\lambda_0(1-b_0^2)+2\sum_{n}\log\left[i\omega_n-\tilde{\varepsilon}_d-\tilde{\Sigma}_d(i\omega_n)\right]\right\}. \label{eq13}
\end{equation} 
The free energy is given by, 
\begin{align}
	\mathcal{F}&=-\frac{1}{\beta}\log(\mathcal{Z}) \nonumber \\
	&=\lambda_0(b_0^2-1)-\frac{2}{\beta}\sum_{n}\log\left[i\omega-\tilde{\varepsilon}-\tilde{\Sigma}(i\omega_n)\right], \label{eq14}
\end{align} 
where the renormalized self energy is, 
\begin{equation}
	\tilde{\Sigma}(i\omega_n)=-\frac{i~\tilde{\Gamma}~\text{sgn}[\text{Im}(i\omega_n)]}{\sqrt{2(i\omega_n+E_F)+E_{\gamma}}}, ~~~ \tilde{\Gamma}=b_0^2\Gamma. \label{eq15}
\end{equation}
The parameters $\lambda_0$ and $b_0^2$ are determined by minimizing the free energy $\mathcal{F}$. We get the following expressions,
\begin{align}
	\frac{2}{\beta}\sum_{n}\frac{1}{i\omega-\tilde{\varepsilon}_d-\tilde{\Sigma}(i\omega_n)}+b_0^2-1&=0 \label{eq16}, \\
	\frac{2}{\beta}\sum_{n}\frac{\tilde{\Sigma}(i\omega)}{i\omega_n-\tilde{\varepsilon}_d-\tilde{\Sigma}(i\omega_n)}+\lambda_0b_0^2&=0. \label{eq17}
\end{align} 
Taking the analytical continuation, $i\omega_n\rightarrow\omega+i0^{+}$, we can rewrite these expressions in the real-frequency formalism, 
\begin{align}
2\int_{-\infty}^{\infty}\mathrm{d}\omega\frac{\rho_d(\omega)}{e^{\beta\omega}+1}+b_0^2-1&=0 \label{eq18}, \\
2\int_{-\infty}^{\infty}\mathrm{d}\omega\frac{(\omega-\tilde{\varepsilon}_d)\rho_d(\omega)}{e^{\beta\omega}+1}+\lambda_0b_0^2&=0. \label{eq19}
\end{align}
where $\rho_d(\omega)$ is the density of impurity states,
\begin{equation}
	\rho_d(\omega)=-\frac{1}{\pi}\frac{\text{Im}~\tilde{\Sigma}(\omega)}{[\omega-\tilde{\varepsilon}-\text{Re}~\tilde{\Sigma}(\omega)]^2+[\text{Im}~\tilde{\Sigma}(\omega)]^2},   \label{eq20}
\end{equation}
$\tilde{\Sigma}(\omega)$ is the analytical continuation of $\tilde{\Sigma}(i\omega_n)$, its explicit expression is given by, 
\begin{align}
	\text{Re}~\tilde{\Sigma}(\omega)=-\frac{\tilde{\Gamma}\Theta(-\omega-E_F-\frac{1}{2}E_{\gamma})}{\sqrt{|2(\omega+E_F)+E_{\gamma}|}},  \label{eq21}\\
	\text{Im}~\tilde{\Sigma}(\omega)=-\frac{\tilde{\Gamma}\Theta(\omega+E_F+\frac{1}{2}E_{\gamma})}{\sqrt{2(\omega+E_F)+E_{\gamma}}},   \label{eq22}
\end{align}
where $\Theta(x)$ is the step function. The Kondo temperature $T_K$ is estimated as the $T$ that makes the slave-boson parameter $b_0^2$ vanishing \cite{PhysRevB.66.045112,NUNES2006313}, the condensation temperature of the slave-boson. 

It is difficult to get the analytical solution of Eqs. (\ref{eq18})-(\ref{eq19}), here we show the numerical results. Unless otherwise noted, in the following calculations we choose the Fermi energy $E_F=0.2$, impurity energy $\varepsilon_d=-0.5$, and hybridization strength $\Gamma=0.05\pi$.  
Fig. \ref{fig4}(a) shows the zero-temperature density of states of the impurity, $\rho_d(\omega)$, calculated using the SBMF method. We find that, in addition to the Kondo resonance peak near the Fermi surface ($\omega=0$), there is another sharp peak. This additional peak is induced by the divergence of the density of states at the conduction band edge. One can find that, when the spin-orbit energy increases from $E_{\gamma}=0$ to $E_{\gamma}=0.7$, the sharp peak come from the conduction band edge is dramatically decreased. Meanwhile, the Kondo resonance peak gets more sharp, which demonstrates that the Kondo temperature is suppressed by the SOC. In order to get a more quantitative result, Fig. \ref{fig4}(b) shows $b_0^2$ as a function of the temperature $T$ for several values of spin-orbit energies varying from $E_{\gamma}=0$ to $E_{\gamma}=0.7$. We find that $b_0^2$ is vanishing in the temperature region $0.01$ to $0.1$, and the Kondo temperature, $T_K$, which is defined as the condensation temperature where $b_0^2=0$, is suppressed by the spin-orbit energy. Quantitatively, Fig. \ref{fig4}(c) shows the variation of $T_K$ vs spin-orbit energy $E_{\gamma}$. One can find that the Kondo temperature is almost a linear function of the spin-orbit energy, and that $T_K$ is suppressed by the SOC. These results are consistent with the estimation from the HFQMC simulations.

\section{Discussion}\label{sec5}
In this paper, we use two different methods, the unbiased HFQMC simulation and the SBMF analysis to study the effect of SOC on the Kondo temperature in a quantum wire. Both the two methods show that the Kondo temperature is suppressed by the SOC, and that the Kondo temperature is almost a linear function of the spin-orbit energy. Our results are qualitatively different from the perturbative renormalization group analysis. In the perturbative renormalization group analysis, the Kondo temperature is always enhanced by the SOC, and an exponential enhancement is predicted. 

For the linear dependence of $T_K$ on the spin-orbit coupling, our results are consistent with Refs. [\onlinecite{PhysRevB.84.193411}] and [\onlinecite{jpcm2016chen}]. However, both Refs. [\onlinecite{PhysRevB.84.193411}] and [\onlinecite{jpcm2016chen}] show that the $T_K$ can be enhanced or depressed by SOC in the two-dimensional electron gas system. For the one-dimensional quantum wire, we find that the Kondo temperature is always suppressed by SOC. This distinction comes from the explicit difference of the form of DOS in the two-dimensional electron gas and the one-dimensional quantum wire. For the two-dimensional electron gas, as shown in [\onlinecite{PhysRevB.84.193411}] and [\onlinecite{jpcm2016chen}], the DOS on the Fermi surface is invariant for different SOCs if $E_F>0$. Which makes the details of $\varepsilon_d$, $U$, and $\Gamma$ being important, and the Kondo temperature being either enhanced or suppressed by SOC. However, for the one-dimensional quantum wire, the DOS on the Fermi surface is variable for different $E_{\gamma}$s. Further more, one can find from Fig. \ref{fig3}(b) that the DOS on the Fermi surface is a decreasing function of $E_{\gamma}$. This influence dominates that the Kondo temperature is suppressed by the SOC.

\section*{Acknowledgment}\label{sec6}
LC would like to express his gratitude to Prof. H.-Q. Lin, J. Sun, and H.-K. Tang for the original HFQMC code. We appreciate the support from the NSFC under Grant Nos. 11504106 and 11447167 and the Fundamental Research Funds for the Central Universities. 

% Specify following sections are appendices. Use \appendix* if there only one appendix.  

%\appendix

%-----------------------------------------------------------------
% Sec**: References
%-----------------------------------------------------------------
%\nocite{*}
\bibliographystyle{apsrev4-1}
\bibliography{G:/5_文章/jabref/magImp}

\begin{thebibliography}{36}%
\makeatletter
\providecommand \@ifxundefined [1]{%
 \@ifx{#1\undefined}
}%
\providecommand \@ifnum [1]{%
 \ifnum #1\expandafter \@firstoftwo
 \else \expandafter \@secondoftwo
 \fi
}%
\providecommand \@ifx [1]{%
 \ifx #1\expandafter \@firstoftwo
 \else \expandafter \@secondoftwo
 \fi
}%
\providecommand \natexlab [1]{#1}%
\providecommand \enquote  [1]{``#1''}%
\providecommand \bibnamefont  [1]{#1}%
\providecommand \bibfnamefont [1]{#1}%
\providecommand \citenamefont [1]{#1}%
\providecommand \href@noop [0]{\@secondoftwo}%
\providecommand \href [0]{\begingroup \@sanitize@url \@href}%
\providecommand \@href[1]{\@@startlink{#1}\@@href}%
\providecommand \@@href[1]{\endgroup#1\@@endlink}%
\providecommand \@sanitize@url [0]{\catcode `\\12\catcode `\$12\catcode
  `\&12\catcode `\#12\catcode `\^12\catcode `\_12\catcode `\%12\relax}%
\providecommand \@@startlink[1]{}%
\providecommand \@@endlink[0]{}%
\providecommand \url  [0]{\begingroup\@sanitize@url \@url }%
\providecommand \@url [1]{\endgroup\@href {#1}{\urlprefix }}%
\providecommand \urlprefix  [0]{URL }%
\providecommand \Eprint [0]{\href }%
\providecommand \doibase [0]{http://dx.doi.org/}%
\providecommand \selectlanguage [0]{\@gobble}%
\providecommand \bibinfo  [0]{\@secondoftwo}%
\providecommand \bibfield  [0]{\@secondoftwo}%
\providecommand \translation [1]{[#1]}%
\providecommand \BibitemOpen [0]{}%
\providecommand \bibitemStop [0]{}%
\providecommand \bibitemNoStop [0]{.\EOS\space}%
\providecommand \EOS [0]{\spacefactor3000\relax}%
\providecommand \BibitemShut  [1]{\csname bibitem#1\endcsname}%
\let\auto@bib@innerbib\@empty
%</preamble>
\bibitem [{\citenamefont {Winkler}(2003)}]{RolandWinkler2003}%
  \BibitemOpen
  \bibfield  {author} {\bibinfo {author} {\bibfnamefont {R.}~\bibnamefont
  {Winkler}},\ }\href@noop {} {\emph {\bibinfo {title} {Spin-orbit Coupling
  Effects in Two-Dimensional Electron and Hole Systems}}}\ (\bibinfo
  {publisher} {Springer-Verlag, Berlin, Heidelberg},\ \bibinfo {year}
  {2003})\BibitemShut {NoStop}%
\bibitem [{\citenamefont {\ifmmode \check{Z}\else
  \v{Z}\fi{}uti\ifmmode~\acute{c}\else \'{c}\fi{}}\ \emph
  {et~al.}(2004)\citenamefont {\ifmmode \check{Z}\else
  \v{Z}\fi{}uti\ifmmode~\acute{c}\else \'{c}\fi{}}, \citenamefont {Fabian},\
  and\ \citenamefont {Das~Sarma}}]{RevModPhys.76.323}%
  \BibitemOpen
  \bibfield  {author} {\bibinfo {author} {\bibfnamefont {I.}~\bibnamefont
  {\ifmmode \check{Z}\else \v{Z}\fi{}uti\ifmmode~\acute{c}\else \'{c}\fi{}}},
  \bibinfo {author} {\bibfnamefont {J.}~\bibnamefont {Fabian}}, \ and\ \bibinfo
  {author} {\bibfnamefont {S.}~\bibnamefont {Das~Sarma}},\ }\href {\doibase
  10.1103/RevModPhys.76.323} {\bibfield  {journal} {\bibinfo  {journal} {Rev.
  Mod. Phys.}\ }\textbf {\bibinfo {volume} {76}},\ \bibinfo {pages} {323}
  (\bibinfo {year} {2004})}\BibitemShut {NoStop}%
\bibitem [{\citenamefont {Koga}\ \emph {et~al.}(2002)\citenamefont {Koga},
  \citenamefont {Nitta}, \citenamefont {Akazaki},\ and\ \citenamefont
  {Takayanagi}}]{PhysRevLett.89.046801}%
  \BibitemOpen
  \bibfield  {author} {\bibinfo {author} {\bibfnamefont {T.}~\bibnamefont
  {Koga}}, \bibinfo {author} {\bibfnamefont {J.}~\bibnamefont {Nitta}},
  \bibinfo {author} {\bibfnamefont {T.}~\bibnamefont {Akazaki}}, \ and\
  \bibinfo {author} {\bibfnamefont {H.}~\bibnamefont {Takayanagi}},\ }\href
  {\doibase 10.1103/PhysRevLett.89.046801} {\bibfield  {journal} {\bibinfo
  {journal} {Phys. Rev. Lett.}\ }\textbf {\bibinfo {volume} {89}},\ \bibinfo
  {pages} {046801} (\bibinfo {year} {2002})}\BibitemShut {NoStop}%
\bibitem [{\citenamefont {Zhu}\ \emph {et~al.}(2011)\citenamefont {Zhu},
  \citenamefont {Cheng},\ and\ \citenamefont
  {Schwingenschl\"ogl}}]{PhysRevB.84.153402}%
  \BibitemOpen
  \bibfield  {author} {\bibinfo {author} {\bibfnamefont {Z.~Y.}\ \bibnamefont
  {Zhu}}, \bibinfo {author} {\bibfnamefont {Y.~C.}\ \bibnamefont {Cheng}}, \
  and\ \bibinfo {author} {\bibfnamefont {U.}~\bibnamefont
  {Schwingenschl\"ogl}},\ }\href {\doibase 10.1103/PhysRevB.84.153402}
  {\bibfield  {journal} {\bibinfo  {journal} {Phys. Rev. B}\ }\textbf {\bibinfo
  {volume} {84}},\ \bibinfo {pages} {153402} (\bibinfo {year}
  {2011})}\BibitemShut {NoStop}%
\bibitem [{\citenamefont {Hasan}\ and\ \citenamefont
  {Kane}(2010)}]{RevModPhys.82.3045}%
  \BibitemOpen
  \bibfield  {author} {\bibinfo {author} {\bibfnamefont {M.~Z.}\ \bibnamefont
  {Hasan}}\ and\ \bibinfo {author} {\bibfnamefont {C.~L.}\ \bibnamefont
  {Kane}},\ }\href {\doibase 10.1103/RevModPhys.82.3045} {\bibfield  {journal}
  {\bibinfo  {journal} {Rev. Mod. Phys.}\ }\textbf {\bibinfo {volume} {82}},\
  \bibinfo {pages} {3045} (\bibinfo {year} {2010})}\BibitemShut {NoStop}%
\bibitem [{\citenamefont {Qi}\ and\ \citenamefont
  {Zhang}(2011)}]{RevModPhys.83.1057}%
  \BibitemOpen
  \bibfield  {author} {\bibinfo {author} {\bibfnamefont {X.-L.}\ \bibnamefont
  {Qi}}\ and\ \bibinfo {author} {\bibfnamefont {S.-C.}\ \bibnamefont {Zhang}},\
  }\href {\doibase 10.1103/RevModPhys.83.1057} {\bibfield  {journal} {\bibinfo
  {journal} {Rev. Mod. Phys.}\ }\textbf {\bibinfo {volume} {83}},\ \bibinfo
  {pages} {1057} (\bibinfo {year} {2011})}\BibitemShut {NoStop}%
\bibitem [{\citenamefont {Lutchyn}\ \emph {et~al.}(2010)\citenamefont
  {Lutchyn}, \citenamefont {Sau},\ and\ \citenamefont
  {Das~Sarma}}]{PhysRevLett.105.077001}%
  \BibitemOpen
  \bibfield  {author} {\bibinfo {author} {\bibfnamefont {R.~M.}\ \bibnamefont
  {Lutchyn}}, \bibinfo {author} {\bibfnamefont {J.~D.}\ \bibnamefont {Sau}}, \
  and\ \bibinfo {author} {\bibfnamefont {S.}~\bibnamefont {Das~Sarma}},\ }\href
  {\doibase 10.1103/PhysRevLett.105.077001} {\bibfield  {journal} {\bibinfo
  {journal} {Phys. Rev. Lett.}\ }\textbf {\bibinfo {volume} {105}},\ \bibinfo
  {pages} {077001} (\bibinfo {year} {2010})}\BibitemShut {NoStop}%
\bibitem [{\citenamefont {Chang}\ \emph {et~al.}(2013)\citenamefont {Chang},
  \citenamefont {Zhang}, \citenamefont {Feng}, \citenamefont {Shen},
  \citenamefont {Zhang}, \citenamefont {Guo}, \citenamefont {Li}, \citenamefont
  {Ou}, \citenamefont {Wei}, \citenamefont {Wang}, \citenamefont {Ji},
  \citenamefont {Feng}, \citenamefont {Ji}, \citenamefont {Chen}, \citenamefont
  {Jia}, \citenamefont {Dai}, \citenamefont {Fang}, \citenamefont {Zhang},
  \citenamefont {He}, \citenamefont {Wang}, \citenamefont {Lu}, \citenamefont
  {Ma},\ and\ \citenamefont {Xue}}]{Chang167}%
  \BibitemOpen
  \bibfield  {author} {\bibinfo {author} {\bibfnamefont {C.-Z.}\ \bibnamefont
  {Chang}}, \bibinfo {author} {\bibfnamefont {J.}~\bibnamefont {Zhang}},
  \bibinfo {author} {\bibfnamefont {X.}~\bibnamefont {Feng}}, \bibinfo {author}
  {\bibfnamefont {J.}~\bibnamefont {Shen}}, \bibinfo {author} {\bibfnamefont
  {Z.}~\bibnamefont {Zhang}}, \bibinfo {author} {\bibfnamefont
  {M.}~\bibnamefont {Guo}}, \bibinfo {author} {\bibfnamefont {K.}~\bibnamefont
  {Li}}, \bibinfo {author} {\bibfnamefont {Y.}~\bibnamefont {Ou}}, \bibinfo
  {author} {\bibfnamefont {P.}~\bibnamefont {Wei}}, \bibinfo {author}
  {\bibfnamefont {L.-L.}\ \bibnamefont {Wang}}, \bibinfo {author}
  {\bibfnamefont {Z.-Q.}\ \bibnamefont {Ji}}, \bibinfo {author} {\bibfnamefont
  {Y.}~\bibnamefont {Feng}}, \bibinfo {author} {\bibfnamefont {S.}~\bibnamefont
  {Ji}}, \bibinfo {author} {\bibfnamefont {X.}~\bibnamefont {Chen}}, \bibinfo
  {author} {\bibfnamefont {J.}~\bibnamefont {Jia}}, \bibinfo {author}
  {\bibfnamefont {X.}~\bibnamefont {Dai}}, \bibinfo {author} {\bibfnamefont
  {Z.}~\bibnamefont {Fang}}, \bibinfo {author} {\bibfnamefont {S.-C.}\
  \bibnamefont {Zhang}}, \bibinfo {author} {\bibfnamefont {K.}~\bibnamefont
  {He}}, \bibinfo {author} {\bibfnamefont {Y.}~\bibnamefont {Wang}}, \bibinfo
  {author} {\bibfnamefont {L.}~\bibnamefont {Lu}}, \bibinfo {author}
  {\bibfnamefont {X.-C.}\ \bibnamefont {Ma}}, \ and\ \bibinfo {author}
  {\bibfnamefont {Q.-K.}\ \bibnamefont {Xue}},\ }\href {\doibase
  10.1126/science.1234414} {\bibfield  {journal} {\bibinfo  {journal}
  {Science}\ }\textbf {\bibinfo {volume} {340}},\ \bibinfo {pages} {167}
  (\bibinfo {year} {2013})}\BibitemShut {NoStop}%
\bibitem [{\citenamefont {Wan}\ \emph {et~al.}(2011)\citenamefont {Wan},
  \citenamefont {Turner}, \citenamefont {Vishwanath},\ and\ \citenamefont
  {Savrasov}}]{PhysRevB.83.205101}%
  \BibitemOpen
  \bibfield  {author} {\bibinfo {author} {\bibfnamefont {X.}~\bibnamefont
  {Wan}}, \bibinfo {author} {\bibfnamefont {A.~M.}\ \bibnamefont {Turner}},
  \bibinfo {author} {\bibfnamefont {A.}~\bibnamefont {Vishwanath}}, \ and\
  \bibinfo {author} {\bibfnamefont {S.~Y.}\ \bibnamefont {Savrasov}},\ }\href
  {\doibase 10.1103/PhysRevB.83.205101} {\bibfield  {journal} {\bibinfo
  {journal} {Phys. Rev. B}\ }\textbf {\bibinfo {volume} {83}},\ \bibinfo
  {pages} {205101} (\bibinfo {year} {2011})}\BibitemShut {NoStop}%
\bibitem [{\citenamefont {Mourik}\ \emph {et~al.}(2012)\citenamefont {Mourik},
  \citenamefont {Zuo}, \citenamefont {Frolov}, \citenamefont {Plissard},
  \citenamefont {Bakkers},\ and\ \citenamefont {Kouwenhoven}}]{Mourik1003}%
  \BibitemOpen
  \bibfield  {author} {\bibinfo {author} {\bibfnamefont {V.}~\bibnamefont
  {Mourik}}, \bibinfo {author} {\bibfnamefont {K.}~\bibnamefont {Zuo}},
  \bibinfo {author} {\bibfnamefont {S.~M.}\ \bibnamefont {Frolov}}, \bibinfo
  {author} {\bibfnamefont {S.~R.}\ \bibnamefont {Plissard}}, \bibinfo {author}
  {\bibfnamefont {E.~P. A.~M.}\ \bibnamefont {Bakkers}}, \ and\ \bibinfo
  {author} {\bibfnamefont {L.~P.}\ \bibnamefont {Kouwenhoven}},\ }\href
  {\doibase 10.1126/science.1222360} {\bibfield  {journal} {\bibinfo  {journal}
  {Science}\ }\textbf {\bibinfo {volume} {336}},\ \bibinfo {pages} {1003}
  (\bibinfo {year} {2012})}\BibitemShut {NoStop}%
\bibitem [{\citenamefont {Das}\ \emph {et~al.}(2012)\citenamefont {Das},
  \citenamefont {Ronen}, \citenamefont {Most}, \citenamefont {Oreg},
  \citenamefont {Heiblum},\ and\ \citenamefont {Shtrikman}}]{Das2012}%
  \BibitemOpen
  \bibfield  {author} {\bibinfo {author} {\bibfnamefont {A.}~\bibnamefont
  {Das}}, \bibinfo {author} {\bibfnamefont {Y.}~\bibnamefont {Ronen}}, \bibinfo
  {author} {\bibfnamefont {Y.}~\bibnamefont {Most}}, \bibinfo {author}
  {\bibfnamefont {Y.}~\bibnamefont {Oreg}}, \bibinfo {author} {\bibfnamefont
  {M.}~\bibnamefont {Heiblum}}, \ and\ \bibinfo {author} {\bibfnamefont
  {H.}~\bibnamefont {Shtrikman}},\ }\href {\doibase 10.1038/nphys2479}
  {\bibfield  {journal} {\bibinfo  {journal} {Nat Phys}\ }\textbf {\bibinfo
  {volume} {8}},\ \bibinfo {pages} {887} (\bibinfo {year} {2012})}\BibitemShut
  {NoStop}%
\bibitem [{\citenamefont {L\"u}\ \emph {et~al.}(2013)\citenamefont {L\"u},
  \citenamefont {Lu}, \citenamefont {Shen},\ and\ \citenamefont
  {Ng}}]{PhysRevB.87.195122}%
  \BibitemOpen
  \bibfield  {author} {\bibinfo {author} {\bibfnamefont {H.-F.}\ \bibnamefont
  {L\"u}}, \bibinfo {author} {\bibfnamefont {H.-Z.}\ \bibnamefont {Lu}},
  \bibinfo {author} {\bibfnamefont {S.-Q.}\ \bibnamefont {Shen}}, \ and\
  \bibinfo {author} {\bibfnamefont {T.-K.}\ \bibnamefont {Ng}},\ }\href
  {\doibase 10.1103/PhysRevB.87.195122} {\bibfield  {journal} {\bibinfo
  {journal} {Phys. Rev. B}\ }\textbf {\bibinfo {volume} {87}},\ \bibinfo
  {pages} {195122} (\bibinfo {year} {2013})}\BibitemShut {NoStop}%
\bibitem [{\citenamefont {Cha}\ \emph {et~al.}(2010)\citenamefont {Cha},
  \citenamefont {Williams}, \citenamefont {Kong}, \citenamefont {Meister},
  \citenamefont {Peng}, \citenamefont {Bestwick}, \citenamefont {Gallagher},
  \citenamefont {Goldhaber-Gordon},\ and\ \citenamefont
  {Cui}}]{nanoLett2010Cha}%
  \BibitemOpen
  \bibfield  {author} {\bibinfo {author} {\bibfnamefont {J.~J.}\ \bibnamefont
  {Cha}}, \bibinfo {author} {\bibfnamefont {J.~R.}\ \bibnamefont {Williams}},
  \bibinfo {author} {\bibfnamefont {D.}~\bibnamefont {Kong}}, \bibinfo {author}
  {\bibfnamefont {S.}~\bibnamefont {Meister}}, \bibinfo {author} {\bibfnamefont
  {H.}~\bibnamefont {Peng}}, \bibinfo {author} {\bibfnamefont {A.~J.}\
  \bibnamefont {Bestwick}}, \bibinfo {author} {\bibfnamefont {P.}~\bibnamefont
  {Gallagher}}, \bibinfo {author} {\bibfnamefont {D.}~\bibnamefont
  {Goldhaber-Gordon}}, \ and\ \bibinfo {author} {\bibfnamefont
  {Y.}~\bibnamefont {Cui}},\ }\href {\doibase 10.1021/nl100146n} {\bibfield
  {journal} {\bibinfo  {journal} {Nano Letters}\ }\textbf {\bibinfo {volume}
  {10}},\ \bibinfo {pages} {1076} (\bibinfo {year} {2010})}\BibitemShut
  {NoStop}%
\bibitem [{\citenamefont {Malecki}(2007)}]{Malecki2007}%
  \BibitemOpen
  \bibfield  {author} {\bibinfo {author} {\bibfnamefont {J.}~\bibnamefont
  {Malecki}},\ }\href {\doibase 10.1007/s10955-007-9414-x} {\bibfield
  {journal} {\bibinfo  {journal} {J. Stat. Phys.}\ }\textbf {\bibinfo {volume}
  {129}},\ \bibinfo {pages} {741} (\bibinfo {year} {2007})}\BibitemShut
  {NoStop}%
\bibitem [{\citenamefont {Zarea}\ \emph {et~al.}(2012)\citenamefont {Zarea},
  \citenamefont {Ulloa},\ and\ \citenamefont
  {Sandler}}]{PhysRevLett.108.046601}%
  \BibitemOpen
  \bibfield  {author} {\bibinfo {author} {\bibfnamefont {M.}~\bibnamefont
  {Zarea}}, \bibinfo {author} {\bibfnamefont {S.~E.}\ \bibnamefont {Ulloa}}, \
  and\ \bibinfo {author} {\bibfnamefont {N.}~\bibnamefont {Sandler}},\ }\href
  {\doibase 10.1103/PhysRevLett.108.046601} {\bibfield  {journal} {\bibinfo
  {journal} {Phys. Rev. Lett.}\ }\textbf {\bibinfo {volume} {108}},\ \bibinfo
  {pages} {046601} (\bibinfo {year} {2012})}\BibitemShut {NoStop}%
\bibitem [{\citenamefont {\ifmmode~\check{Z}\else \v{Z}\fi{}itko}\ and\
  \citenamefont {Bon\ifmmode~\check{c}\else
  \v{c}\fi{}a}(2011)}]{PhysRevB.84.193411}%
  \BibitemOpen
  \bibfield  {author} {\bibinfo {author} {\bibfnamefont {R.}~\bibnamefont
  {\ifmmode~\check{Z}\else \v{Z}\fi{}itko}}\ and\ \bibinfo {author}
  {\bibfnamefont {J.}~\bibnamefont {Bon\ifmmode~\check{c}\else \v{c}\fi{}a}},\
  }\href {\doibase 10.1103/PhysRevB.84.193411} {\bibfield  {journal} {\bibinfo
  {journal} {Phys. Rev. B}\ }\textbf {\bibinfo {volume} {84}},\ \bibinfo
  {pages} {193411} (\bibinfo {year} {2011})}\BibitemShut {NoStop}%
\bibitem [{\citenamefont {Isaev}\ \emph {et~al.}(2012)\citenamefont {Isaev},
  \citenamefont {Agterberg},\ and\ \citenamefont
  {Vekhter}}]{PhysRevB.85.081107}%
  \BibitemOpen
  \bibfield  {author} {\bibinfo {author} {\bibfnamefont {L.}~\bibnamefont
  {Isaev}}, \bibinfo {author} {\bibfnamefont {D.~F.}\ \bibnamefont
  {Agterberg}}, \ and\ \bibinfo {author} {\bibfnamefont {I.}~\bibnamefont
  {Vekhter}},\ }\href {\doibase 10.1103/PhysRevB.85.081107} {\bibfield
  {journal} {\bibinfo  {journal} {Phys. Rev. B}\ }\textbf {\bibinfo {volume}
  {85}},\ \bibinfo {pages} {081107} (\bibinfo {year} {2012})}\BibitemShut
  {NoStop}%
\bibitem [{\citenamefont {Wong}\ \emph {et~al.}(2016)\citenamefont {Wong},
  \citenamefont {Ulloa}, \citenamefont {Sandler},\ and\ \citenamefont
  {Ingersent}}]{PhysRevB.93.075148}%
  \BibitemOpen
  \bibfield  {author} {\bibinfo {author} {\bibfnamefont {A.}~\bibnamefont
  {Wong}}, \bibinfo {author} {\bibfnamefont {S.~E.}\ \bibnamefont {Ulloa}},
  \bibinfo {author} {\bibfnamefont {N.}~\bibnamefont {Sandler}}, \ and\
  \bibinfo {author} {\bibfnamefont {K.}~\bibnamefont {Ingersent}},\ }\href
  {\doibase 10.1103/PhysRevB.93.075148} {\bibfield  {journal} {\bibinfo
  {journal} {Phys. Rev. B}\ }\textbf {\bibinfo {volume} {93}},\ \bibinfo
  {pages} {075148} (\bibinfo {year} {2016})}\BibitemShut {NoStop}%
\bibitem [{\citenamefont {Agarwala}\ and\ \citenamefont
  {Shenoy}(2016)}]{PhysRevB.93.241111}%
  \BibitemOpen
  \bibfield  {author} {\bibinfo {author} {\bibfnamefont {A.}~\bibnamefont
  {Agarwala}}\ and\ \bibinfo {author} {\bibfnamefont {V.~B.}\ \bibnamefont
  {Shenoy}},\ }\href {\doibase 10.1103/PhysRevB.93.241111} {\bibfield
  {journal} {\bibinfo  {journal} {Phys. Rev. B}\ }\textbf {\bibinfo {volume}
  {93}},\ \bibinfo {pages} {241111} (\bibinfo {year} {2016})}\BibitemShut
  {NoStop}%
\bibitem [{\citenamefont {Chen}\ \emph {et~al.}(2016)\citenamefont {Chen},
  \citenamefont {Sun}, \citenamefont {Tang},\ and\ \citenamefont
  {Lin}}]{jpcm2016chen}%
  \BibitemOpen
  \bibfield  {author} {\bibinfo {author} {\bibfnamefont {L.}~\bibnamefont
  {Chen}}, \bibinfo {author} {\bibfnamefont {J.}~\bibnamefont {Sun}}, \bibinfo
  {author} {\bibfnamefont {H.-K.}\ \bibnamefont {Tang}}, \ and\ \bibinfo
  {author} {\bibfnamefont {H.-Q.}\ \bibnamefont {Lin}},\ }\href
  {http://stacks.iop.org/0953-8984/28/i=39/a=396005} {\bibfield  {journal}
  {\bibinfo  {journal} {J. Phys.: Condens. Matter}\ }\textbf {\bibinfo {volume}
  {28}},\ \bibinfo {pages} {396005} (\bibinfo {year} {2016})}\BibitemShut
  {NoStop}%
\bibitem [{\citenamefont {de~Sousa}\ \emph {et~al.}(2016)\citenamefont
  {de~Sousa}, \citenamefont {Silva},\ and\ \citenamefont
  {Vernek}}]{PhysRevB.94.125115}%
  \BibitemOpen
  \bibfield  {author} {\bibinfo {author} {\bibfnamefont {G.~R.}\ \bibnamefont
  {de~Sousa}}, \bibinfo {author} {\bibfnamefont {J.~F.}\ \bibnamefont {Silva}},
  \ and\ \bibinfo {author} {\bibfnamefont {E.}~\bibnamefont {Vernek}},\ }\href
  {\doibase 10.1103/PhysRevB.94.125115} {\bibfield  {journal} {\bibinfo
  {journal} {Phys. Rev. B}\ }\textbf {\bibinfo {volume} {94}},\ \bibinfo
  {pages} {125115} (\bibinfo {year} {2016})}\BibitemShut {NoStop}%
\bibitem [{\citenamefont {Hirsch}\ and\ \citenamefont
  {Fye}(1986)}]{PhysRevLett.56.2521}%
  \BibitemOpen
  \bibfield  {author} {\bibinfo {author} {\bibfnamefont {J.~E.}\ \bibnamefont
  {Hirsch}}\ and\ \bibinfo {author} {\bibfnamefont {R.~M.}\ \bibnamefont
  {Fye}},\ }\href {\doibase 10.1103/PhysRevLett.56.2521} {\bibfield  {journal}
  {\bibinfo  {journal} {Phys. Rev. Lett.}\ }\textbf {\bibinfo {volume} {56}},\
  \bibinfo {pages} {2521} (\bibinfo {year} {1986})}\BibitemShut {NoStop}%
\bibitem [{\citenamefont {Fye}\ and\ \citenamefont
  {Hirsch}(1988)}]{PhysRevB.38.433}%
  \BibitemOpen
  \bibfield  {author} {\bibinfo {author} {\bibfnamefont {R.~M.}\ \bibnamefont
  {Fye}}\ and\ \bibinfo {author} {\bibfnamefont {J.~E.}\ \bibnamefont
  {Hirsch}},\ }\href {\doibase 10.1103/PhysRevB.38.433} {\bibfield  {journal}
  {\bibinfo  {journal} {Phys. Rev. B}\ }\textbf {\bibinfo {volume} {38}},\
  \bibinfo {pages} {433} (\bibinfo {year} {1988})}\BibitemShut {NoStop}%
\bibitem [{\citenamefont {Fye}\ \emph {et~al.}(1987)\citenamefont {Fye},
  \citenamefont {Hirsch},\ and\ \citenamefont {Scalapino}}]{PhysRevB.35.4901}%
  \BibitemOpen
  \bibfield  {author} {\bibinfo {author} {\bibfnamefont {R.~M.}\ \bibnamefont
  {Fye}}, \bibinfo {author} {\bibfnamefont {J.~E.}\ \bibnamefont {Hirsch}}, \
  and\ \bibinfo {author} {\bibfnamefont {D.~J.}\ \bibnamefont {Scalapino}},\
  }\href {\doibase 10.1103/PhysRevB.35.4901} {\bibfield  {journal} {\bibinfo
  {journal} {Phys. Rev. B}\ }\textbf {\bibinfo {volume} {35}},\ \bibinfo
  {pages} {4901} (\bibinfo {year} {1987})}\BibitemShut {NoStop}%
\bibitem [{\citenamefont {Fye}\ and\ \citenamefont
  {Hirsch}(1989)}]{PhysRevB.40.4780}%
  \BibitemOpen
  \bibfield  {author} {\bibinfo {author} {\bibfnamefont {R.~M.}\ \bibnamefont
  {Fye}}\ and\ \bibinfo {author} {\bibfnamefont {J.~E.}\ \bibnamefont
  {Hirsch}},\ }\href {\doibase 10.1103/PhysRevB.40.4780} {\bibfield  {journal}
  {\bibinfo  {journal} {Phys. Rev. B}\ }\textbf {\bibinfo {volume} {40}},\
  \bibinfo {pages} {4780} (\bibinfo {year} {1989})}\BibitemShut {NoStop}%
\bibitem [{\citenamefont {Bulut}\ \emph {et~al.}(2007)\citenamefont {Bulut},
  \citenamefont {Tanikawa}, \citenamefont {Takahashi},\ and\ \citenamefont
  {Maekawa}}]{PhysRevB.76.045220}%
  \BibitemOpen
  \bibfield  {author} {\bibinfo {author} {\bibfnamefont {N.}~\bibnamefont
  {Bulut}}, \bibinfo {author} {\bibfnamefont {K.}~\bibnamefont {Tanikawa}},
  \bibinfo {author} {\bibfnamefont {S.}~\bibnamefont {Takahashi}}, \ and\
  \bibinfo {author} {\bibfnamefont {S.}~\bibnamefont {Maekawa}},\ }\href
  {\doibase 10.1103/PhysRevB.76.045220} {\bibfield  {journal} {\bibinfo
  {journal} {Phys. Rev. B}\ }\textbf {\bibinfo {volume} {76}},\ \bibinfo
  {pages} {045220} (\bibinfo {year} {2007})}\BibitemShut {NoStop}%
\bibitem [{\citenamefont {Hu}\ \emph {et~al.}(2011)\citenamefont {Hu},
  \citenamefont {Ma}, \citenamefont {Lin},\ and\ \citenamefont
  {Gubernatis}}]{PhysRevB.84.075414}%
  \BibitemOpen
  \bibfield  {author} {\bibinfo {author} {\bibfnamefont {F.~M.}\ \bibnamefont
  {Hu}}, \bibinfo {author} {\bibfnamefont {T.}~\bibnamefont {Ma}}, \bibinfo
  {author} {\bibfnamefont {H.-Q.}\ \bibnamefont {Lin}}, \ and\ \bibinfo
  {author} {\bibfnamefont {J.~E.}\ \bibnamefont {Gubernatis}},\ }\href
  {\doibase 10.1103/PhysRevB.84.075414} {\bibfield  {journal} {\bibinfo
  {journal} {Phys. Rev. B}\ }\textbf {\bibinfo {volume} {84}},\ \bibinfo
  {pages} {075414} (\bibinfo {year} {2011})}\BibitemShut {NoStop}%
\bibitem [{\citenamefont {Sun}\ \emph {et~al.}(2014)\citenamefont {Sun},
  \citenamefont {Chen},\ and\ \citenamefont {Lin}}]{PhysRevB.89.115101}%
  \BibitemOpen
  \bibfield  {author} {\bibinfo {author} {\bibfnamefont {J.}~\bibnamefont
  {Sun}}, \bibinfo {author} {\bibfnamefont {L.}~\bibnamefont {Chen}}, \ and\
  \bibinfo {author} {\bibfnamefont {H.-Q.}\ \bibnamefont {Lin}},\ }\href
  {\doibase 10.1103/PhysRevB.89.115101} {\bibfield  {journal} {\bibinfo
  {journal} {Phys. Rev. B}\ }\textbf {\bibinfo {volume} {89}},\ \bibinfo
  {pages} {115101} (\bibinfo {year} {2014})}\BibitemShut {NoStop}%
\bibitem [{\citenamefont {Krishna-murthy}\ \emph {et~al.}(1980)\citenamefont
  {Krishna-murthy}, \citenamefont {Wilkins},\ and\ \citenamefont
  {Wilson}}]{PhysRevB.21.1003}%
  \BibitemOpen
  \bibfield  {author} {\bibinfo {author} {\bibfnamefont {H.~R.}\ \bibnamefont
  {Krishna-murthy}}, \bibinfo {author} {\bibfnamefont {J.~W.}\ \bibnamefont
  {Wilkins}}, \ and\ \bibinfo {author} {\bibfnamefont {K.~G.}\ \bibnamefont
  {Wilson}},\ }\href {\doibase 10.1103/PhysRevB.21.1003} {\bibfield  {journal}
  {\bibinfo  {journal} {Phys. Rev. B}\ }\textbf {\bibinfo {volume} {21}},\
  \bibinfo {pages} {1003} (\bibinfo {year} {1980})}\BibitemShut {NoStop}%
\bibitem [{\citenamefont {Coleman}(1984)}]{PhysRevB.29.3035}%
  \BibitemOpen
  \bibfield  {author} {\bibinfo {author} {\bibfnamefont {P.}~\bibnamefont
  {Coleman}},\ }\href {\doibase 10.1103/PhysRevB.29.3035} {\bibfield  {journal}
  {\bibinfo  {journal} {Phys. Rev. B}\ }\textbf {\bibinfo {volume} {29}},\
  \bibinfo {pages} {3035} (\bibinfo {year} {1984})}\BibitemShut {NoStop}%
\bibitem [{\citenamefont {Coleman}(1987)}]{PhysRevB.35.5072}%
  \BibitemOpen
  \bibfield  {author} {\bibinfo {author} {\bibfnamefont {P.}~\bibnamefont
  {Coleman}},\ }\href {\doibase 10.1103/PhysRevB.35.5072} {\bibfield  {journal}
  {\bibinfo  {journal} {Phys. Rev. B}\ }\textbf {\bibinfo {volume} {35}},\
  \bibinfo {pages} {5072} (\bibinfo {year} {1987})}\BibitemShut {NoStop}%
\bibitem [{\citenamefont {Dorin}\ and\ \citenamefont
  {Schlottmann}(1993)}]{PhysRevB.47.5095}%
  \BibitemOpen
  \bibfield  {author} {\bibinfo {author} {\bibfnamefont {V.}~\bibnamefont
  {Dorin}}\ and\ \bibinfo {author} {\bibfnamefont {P.}~\bibnamefont
  {Schlottmann}},\ }\href {\doibase 10.1103/PhysRevB.47.5095} {\bibfield
  {journal} {\bibinfo  {journal} {Phys. Rev. B}\ }\textbf {\bibinfo {volume}
  {47}},\ \bibinfo {pages} {5095} (\bibinfo {year} {1993})}\BibitemShut
  {NoStop}%
\bibitem [{\citenamefont {Franco}\ \emph {et~al.}(2002)\citenamefont {Franco},
  \citenamefont {Figueira},\ and\ \citenamefont {Foglio}}]{PhysRevB.66.045112}%
  \BibitemOpen
  \bibfield  {author} {\bibinfo {author} {\bibfnamefont {R.}~\bibnamefont
  {Franco}}, \bibinfo {author} {\bibfnamefont {M.~S.}\ \bibnamefont
  {Figueira}}, \ and\ \bibinfo {author} {\bibfnamefont {M.~E.}\ \bibnamefont
  {Foglio}},\ }\href {\doibase 10.1103/PhysRevB.66.045112} {\bibfield
  {journal} {\bibinfo  {journal} {Phys. Rev. B}\ }\textbf {\bibinfo {volume}
  {66}},\ \bibinfo {pages} {045112} (\bibinfo {year} {2002})}\BibitemShut
  {NoStop}%
\bibitem [{\citenamefont {Nunes}\ \emph {et~al.}(2006)\citenamefont {Nunes},
  \citenamefont {Figueira},\ and\ \citenamefont {Foglio}}]{NUNES2006313}%
  \BibitemOpen
  \bibfield  {author} {\bibinfo {author} {\bibfnamefont {L.~H.}\ \bibnamefont
  {Nunes}}, \bibinfo {author} {\bibfnamefont {M.}~\bibnamefont {Figueira}}, \
  and\ \bibinfo {author} {\bibfnamefont {M.}~\bibnamefont {Foglio}},\ }\href
  {\doibase http://dx.doi.org/10.1016/j.physleta.2006.05.025} {\bibfield
  {journal} {\bibinfo  {journal} {Physics Letters A}\ }\textbf {\bibinfo
  {volume} {358}},\ \bibinfo {pages} {313 } (\bibinfo {year}
  {2006})}\BibitemShut {NoStop}%
\bibitem [{\citenamefont {Zhu}\ and\ \citenamefont
  {Ting}(2000)}]{PhysRevB.63.020506}%
  \BibitemOpen
  \bibfield  {author} {\bibinfo {author} {\bibfnamefont {J.-X.}\ \bibnamefont
  {Zhu}}\ and\ \bibinfo {author} {\bibfnamefont {C.~S.}\ \bibnamefont {Ting}},\
  }\href {\doibase 10.1103/PhysRevB.63.020506} {\bibfield  {journal} {\bibinfo
  {journal} {Phys. Rev. B}\ }\textbf {\bibinfo {volume} {63}},\ \bibinfo
  {pages} {020506} (\bibinfo {year} {2000})}\BibitemShut {NoStop}%
\bibitem [{\citenamefont {Akbari}\ \emph {et~al.}(2010)\citenamefont {Akbari},
  \citenamefont {Eremin},\ and\ \citenamefont
  {Thalmeier}}]{PhysRevB.81.014524}%
  \BibitemOpen
  \bibfield  {author} {\bibinfo {author} {\bibfnamefont {A.}~\bibnamefont
  {Akbari}}, \bibinfo {author} {\bibfnamefont {I.}~\bibnamefont {Eremin}}, \
  and\ \bibinfo {author} {\bibfnamefont {P.}~\bibnamefont {Thalmeier}},\ }\href
  {\doibase 10.1103/PhysRevB.81.014524} {\bibfield  {journal} {\bibinfo
  {journal} {Phys. Rev. B}\ }\textbf {\bibinfo {volume} {81}},\ \bibinfo
  {pages} {014524} (\bibinfo {year} {2010})}\BibitemShut {NoStop}%
\end{thebibliography}%

\end{document}